\documentclass[%
showpacs, showkeys,
,preprint%
 ,onecolumn%
 ,secnumarabic%
,amssymb, amsmath,nobibnotes, aps, prl]{revtex4}
\usepackage{docs}%
\usepackage{bm}%
\usepackage{graphicx}
\usepackage{longtable}%
\usepackage{bm}%
\usepackage{docs}

\expandafter\ifx\csname package@font\endcsname\relax\else
 \expandafter\expandafter
 \expandafter\usepackage
 \expandafter\expandafter
 \expandafter{\csname package@font\endcsname}%
\fi

\begin{document}

\title[Modular functions and Ramanujan sums for the analysis \\ of $1/f$ noise in electronic circuits]
{Modular functions and Ramanujan sums for the analysis \\
of $1/f$ noise in electronic circuits
 }%
\author{Michel Planat}
\email{planat@lpmo.edu}
 \affiliation{Laboratoire de Physique et
M\'{e}trologie des Oscillateurs du CNRS\\32 Avenue de
l'Observatoire, 25044 Besan\c{c}on Cedex, France }

\begin{abstract}

\vspace{1.5ex}A number theoretical model of $1/f$ noise found in
phase locked loops is developed. The dynamics of phases and
frequencies involved in the nonlinear mixing of oscillators and
the low-pass filtering is formulated thanks to the rules of the
hyperbolic geometry of the half plane. A  cornerstone of the
analysis is the Ramanujan sums expansion of arithmetical functions
found in prime number theory, and their link to Riemann
hypothesis.
   \end{abstract}

 \maketitle

\section{Introduction}
$1/f$ noise, already discovered by Nyquist in resistors at the
dawn of electronic age, still fascinates experimentalists and
theoreticians. This is because none physical principle or
mathematical function allows to predict easily the observed
hyperbolic $1/f$ power spectral density. Nevertheless the
occurrence of $1/f$ f\mbox{}luctuations in areas as diverse as
electronics, chemistry, biology, cognition or geology claims for
an unifying mathematical principle \cite{WentianLi}.

A newly discovered clue for $1/f$ noise lies in the concept of a
phase locked loop (or PLL) \cite{APL02}. In essence two
interacting oscillators, whatever their origin, attempt to
cooperate by locking their frequency and their phase. They can do
it by exchanging continuously tiny amounts of energy, so that both
the coupling coeff\mbox{}icient and the beat frequency should
f\mbox{}luctuate in time around their average value. Correlations
between amplitude and frequency noise were observed \cite{Yamoto}.

Fortunately one can gain a good level of understanding of phase
locking from quartz crystal oscillators used in high frequency
synthesizers, ultrastable clocks and communication engineering
(mobile phones or so). The PLL used in a FM radio receiver is a
genuine generator of $1/f$ noise. Close to phase locking the level
of $1/f$ noise scales approximately as $\tilde{\sigma}^2$, where
$\tilde{\sigma}=\sigma K/\tilde{\omega_B}$ is the ratio between
the open loop gain $K$ and the beat frequency $\tilde{\omega_B}$
times a constant coeff\mbox{}icient $\sigma$. The relation above
is explained from a simple non linear model of the PLL known as
Adler's equation
\begin{equation}
\dot{\theta}(t)+K H(P)\sin \theta(t)=\omega_B, \label{equation1}
\end{equation}
where at this stage $H(P)=1$, $\omega_B=\omega(t)-\omega_0$ is the
angular frequency shift between the two quartz oscillators at the
input of the non linear device (a Schottky diodes double balanced
mixer), and $\theta(t)$ is the phase shift of the two oscillators
versus time t. Solving (\ref {equation1}) and differentiating one
gets the observed noise level $\tilde{\sigma}$ versus the one
$\sigma=\delta \omega_B/\tilde{\omega_B}$ for the open loop case.
Thus the model doesn't explain the existence of $1/f$ noise but
correctly predicts its dependance on the physical parameters of
the loop \cite{APL02}.

\section{Low Pass Filtering and $1/f$ Noise}
Besides one can get detailed knowledge of harmonic conversions in
the PLL by accounting for the transfer function $H(P)$, where
$P=\frac{d}{dt}$ is the Laplace operator. If $H(P)$ is a low pass
f\mbox{}iltering function with cut-off frequency $f_c$, the
frequency at the output of the mixer + f\mbox{}ilter stage is such
that
\begin{equation}
\mu=f_B(t)/f_0=q_i|\nu-p_i/q_i| \le f_c/f_0,~~
\mbox{$p_i$~and~$q_i$~integers}. \label{equation2}
\end{equation}
The beat frequency $f_B(t)$ results from the continued fraction
expansion of the input frequency ratio
\begin{equation}
\nu=f(t)/f_0=[a_0;a_1,a_2,\ldots a_i,a,\ldots ], \\
\label{equation3}
\end{equation}
where the brackets mean expansions
$a_0+1/(a_1+1/(a_2+1/\ldots+1/(a_i+1/(a+\ldots))))$. The
truncation at the integer part $a=[\frac{f_0}{f_c q_i}]$
def\mbox{}ines the edges of the basin; they are located at
$\nu_1=[a_0;a_1,a_2,\ldots,a_i, a]$ and $\nu_2=[a_0;a_1,a_2,\ldots
a_i-1,1,a]$ \cite{Fluc01}. The two expansions in $\nu_1$ and
$\nu_2$, prior to the last f\mbox{}iltering partial quotient $a$,
are the two allowed ones for a rational number. The convergents
$p_i/q_i$ at level $i$ are obtained using the matrix products
\begin{equation}
\left[\begin{array}{cc} a_0 & 1\\ 1 & 0 \end{array}\right]
\left[\begin{array}{cc} a_1 & 1\\ 1 & 0 \end{array}\right]\cdots
\left[\begin{array}{cc} a_i & 1\\ 1 & 0 \end{array}\right]
=\left[\begin{array}{cc} p_i&p_{i-1}\\ q_i&q_{i-1}
\end{array}\right]. \label{matrices}
\end{equation}
\begin{figure}[htb]
\centerline{
\includegraphics[width=6.5cm]{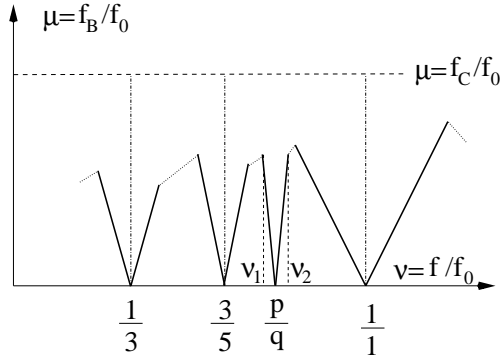}}
\caption{The intermodulation spectrum at the output of the
mixer+filter set-up}
\end{figure}

Using (\ref{matrices}) one can get the fractions $\nu_1$ and
$\nu_2$ as $\nu_1=\frac{p_a}{q_a}$ and
$\nu_2=\frac{p_i(2a+1)-p_a}{q_i(2a+1)-q_a}$ so that, with the
relation relating convergents  $(p_i
q_{i-1}-p_{i-1}q_i)=(-1)^{i-1}$, the width of the basin of index
$i$ is
$|\nu_1-\nu_2|=\frac{2a+1}{q_a(q_a+(2a+1)q_i)}\simeq\frac{1}{q_a
q_i}$ whenever $a>1$ (see also Fig. 1).

 In our previous publications we proposed a phenomenological model
 for $1/f$ noise in the PLL, based on an arithmetical function
 which is a logarithmic coding for prime numbers \cite{APL02},\cite{Fluc01}. If one accepts a
 coupling coeff\mbox{}icient evolving discontinuously versus the time $n$
 as $K=K_0\Lambda(n)$, with $\Lambda(n)$ the Mangoldt function
which is $\ln(p)$ if $n$ is the power of a prime number $p$ and
$0$ otherwise, then the average coupling coeff\mbox{}icient is
$K_0$ and there is an arithmetical f\mbox{}luctuation
$\epsilon(t)$
\begin{eqnarray}
&\psi(t)=\sum_{n=1}^t\Lambda(n)=t(1+\epsilon(t)),\nonumber\\
&t\epsilon(t)=-\ln(2\pi)-\frac{1}{2}\ln(1-t^{-2})-\sum_{\rho}\frac{t^{\rho}}{\rho}.
\label{Riemann}
\end{eqnarray}
The three terms at the right hand side of $t\epsilon(t)$ come from
the singularities of the Riemann zeta function $\zeta(s)$, that
are the pole at $s=1$, the trivial zeros at $s=-2l$, $l$ integer,
and the zeros on the critical line
$\Re(s)=\frac{1}{2}$\cite{Fluc01}. Also the power spectral density
roughly shows a $1/f$ dependance versus the Fourier frequency $f$.
This is an unexpected relation between Riemann zeros (the unsolved
Riemann hypothesis is that all zeros should lie on the critical
line) and $1/f$ noise.

We improved the model by replacing the Mangoldt function by its
modif\mbox{}ied form $b(n)=\Lambda(n)\phi(n)/n$, with $\phi(n)$
the Euler (totient) function \cite{PRE02}. This seemingly
insignif\mbox{}icant change was introduced by Hardy \cite{Hardy}
in the context of Ramanujan sums for the Goldbach conjecture and
resurrected recently by Gadiyar and Padma for analyzing the
distribution of prime pairs \cite{Gadiyar}. Then by defining the
error term $\epsilon_B(t)$ from
\begin{equation}
B(t)=\sum_{n=1}^t b(n)=t(1+\epsilon_B(t)), \label{functionB}
\end{equation}
its power spectral density $S_B(f)\simeq\frac{1}{f^{2\alpha}}$
exhibits a slope close to twice the Golden ratio
$\alpha\simeq(\sqrt5-1)/2\simeq0.618$ (see Fig. 2).

\begin{figure}[htb]
\centerline{
\includegraphics[width=6.5cm]{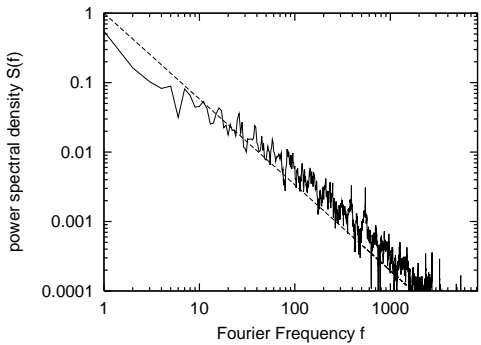}}
\caption{Power spectral density of the error term in
modif\mbox{}ied Mangoldt function b(n) in comparison to the power
law $1/f^{2\alpha}$, with the Golden ratio
$\alpha=(\sqrt{5}-1)/2$.}
\end{figure}
The modif\mbox{}ied Mangoldt function occurs in a natural way from
the logarithmic derivative of the quotient
\begin{equation}
Z(s)=\frac{\zeta(s)}{\zeta(s+1)}=\sum_{n\ge
1}\frac{\phi(n)}{n^{s+1}}, \label{equation11}
\end{equation}
since $-\frac{Z'(s)}{Z(s)}=\sum_{n\ge 1}\frac{b(n)}{n^s}$. This
replaces the relation from the Riemann zeta function where
$-\frac{\zeta'(s)}{\zeta(s)}=\sum_{n\ge 1}\frac{\Lambda(n)}{n^s}$.

\section{The Hyperbolic Geometry of Phase Noise and $1/f$ Frequency Noise}
The whole theory can be justified by studying the noise in the
half plane $\it{H}=\{z=\nu+Iy,~I^2=-1,~y>0\}$ of coordinates
$\nu=\frac{f}{f_0}$ and $ y=\frac{f_B}{f_c}>0$ and by introducing
the modular transformations
\begin{equation}
z \rightarrow \gamma(z)=\frac{p_i z+p'_i}{q_i z+q'_i},
~~p_iq'_i-p'_iq_i=1. \label{equation13}
\end{equation}
The set of images of the f\mbox{}iltering line $z=\nu+I$ under all
modular transformations can be written as
\begin{equation}
|z-(\frac{p_i}{q_i} + \frac{I}{2q_i^2})|=\frac{1}{2q_i^2}.
\label{Ford}
\end{equation}
Equation (\ref{Ford}) def\mbox{}ines Ford circles
\cite{Rademacher}\cite{Planat2}(see Fig. 3) centered at points
$z=\frac{p_i}{q_i}+\frac{I}{2q_i^2}$ with radius
$\frac{1}{2q_i^2}$ \cite{Rademacher}. To each $\frac{p_i}{q_i}$
belongs a Ford circle in the upper half plane, which is tangent to
the real axis at $\nu=\frac{p_i}{q_i}$. Ford circles never
intersect: they are tangent to each other if and only if they
belong to fractions which are adjacent in the Farey sequence
$\frac{0}{1}<\cdots\frac{p_1}{q_1}<\frac{p_1+p_2}{q_1+q_2}<\frac{q_2}{q_2}\cdots<\frac{1}{1}$.
\begin{figure}[htb]
\centerline{
\includegraphics[width=6.5cm]{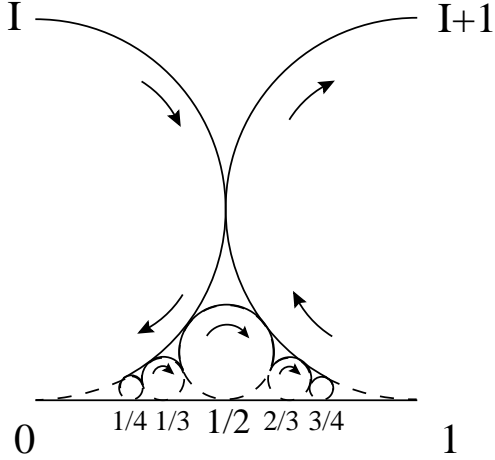}}
\caption{Ford circles: the mapping of the f\mbox{}iltering line
under modular transformations (\ref{equation13}). The arrows
indicates that Ford circles were used as an integration path by
Rademacher to compute the partition function $p(n)$.}
\end{figure}
\begin{figure}[htb]
\centerline{
\includegraphics[width=6.5cm]{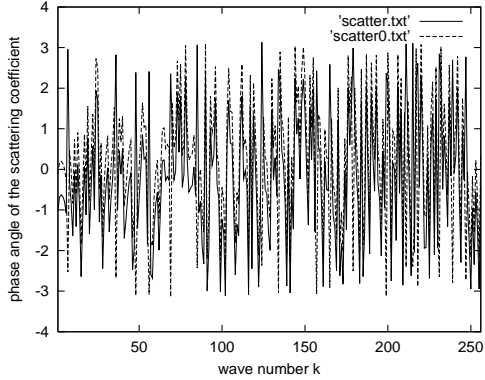}}
\caption{The phase angle $\theta(k)$ for the scattering of noise
waves on the modular surface. }
\end{figure}
%

%

For modular transformations one easily calculates $\Im
\gamma(z)=\frac{y}{|q_iz+q_i'|^2},~\frac{d
\gamma(z)}{dz}=\frac{1}{(q_iz+q_i')^2}$. where the symbol $\Im$
means the imaginary part. This implies $|\frac{d
\gamma(z)}{dz}|(\Im \gamma(z))^{-1}=(\Im z)^{-1}$ and the
invariance of the non-Euclidean metric
$\frac{|dz|}{y}=\frac{(d\nu^2+dy^2)^{1/2}}{y}$.

A basic fact about the modular transformations (\ref{equation13})
is that they form a discontinuous group $\Gamma\simeq
SL(2,\it{Z})/\{\pm1\}$, which is called the modular group. The
action of $\Gamma$ on the half-plane $\it{H}$ looks like the one
generated by two independent linear translations on the Euclidean
plane, which is equivalent to tesselate the complex plane $\it{C}$
with congruent parallelograms. Here one introduces the fundamental
domain of $\Gamma$ (or modular surface) $\it{F}=\{z \in
\it{H}:~|z|\ge 1,~|\nu|\le \frac{1}{2}\}$, and the family of
domains $\{\gamma(\it{F}),\gamma \in \Gamma\}$ induces a
tesselation of $\it{H}$\cite{Serre},\cite{Gutzwiller}.

Corresponding to the Riemannian metric there is the non-Euclidean
Laplacian $\Delta=y^2(\frac{\partial^2}{\partial
\nu^2}+\frac{\partial^2}{\partial y^2})$, and one can write a
non-Euclidean wave equation \cite{Gutzwiller}
\begin{equation}
i\hbar \frac{\partial \Psi}{\partial
t}=-\frac{\hbar^2}{2m}(\Delta+\frac{1}{4})\Psi.
\end{equation}
Stationary solutions $\Psi=\phi(\nu,y) \exp(-I\omega t)$ satisfy
\begin{equation}
(\Delta+\frac{1}{4}+k^2)\phi=0,~~\mbox{with}~k^2=\frac{2mE}{\hbar^2},~E=\hbar
\omega.
\end{equation}
They are special $\nu$-independent solutions
$\phi=y^s~~\mbox{with}~s=\frac{1}{2}+I k$, and any other wave can
be transformed by (\ref{equation13}) to form the new solution
\begin{equation}
\phi_s(z)=(\Im
\gamma(z))^s=\frac{y^s}{|q_iz+q_i'|^{2s}},~s=\frac{1}{2}+I k.
\label{equation24}
\end{equation}
In Gutzwiller's paper \cite{Gutzwiller} there is a detailed
discussion on the geometry of solutions (\ref{equation24}). The
$\nu$-independent solution represents a wave propagating in the
y-direction $\Psi(y)=y^{1/2}\exp (Ik\ln y -I\omega t)$. The factor
$y^{1/2}$ comes from restricting the total f\mbox{}lux in a
vertical strip of constant width such as $0<\nu<1$, so that with
the hyperbolic metric one has $\int_0^1 d \nu/y=1/y$. The phase
factor is explained by looking at wavefronts $y=C^{\rm{te}}$ and
the distance proportional to them is $\int_1^y dy/y=\ln y$.

For equation (\ref{equation24}) the wavefronts are obtained by
assigning to the quantity $\frac{y}{|q_i z+q_i'|^2}$ some constant
value that one can choose to be the unity. These wavefronts are
circles of equation
$(\nu+\frac{q_i'}{q_i})^2+(y-\frac{1}{2q_i^2})^2=\frac{1}{4q_i^4}$.
They are tangent to the real axis at $\nu_0=-\frac{q_i'}{q_i}$ and
of radius $\frac{1}{2q_i^2}$. Up to a constant shift along the
real axis $\nu$, they are the same as the Ford circles shown on
Fig. 3.

Thus an outgoing plane wave can be said to  start at $y=1$, and
move up into $\it{H}$ with increasing $y$, its wavefronts parallel
to the real axis. All other plane waves start at some circle which
touches the real axis at some arbitrary point $\nu_0$ of radius
$\frac{1}{2q_i^2}$. Then they contract by shrinking their radius
while maintaining their point of contact $\nu_0$. These wave
fronts are horocycles in the hyperbolic geometry of the
half-plane. They are perpendicular to the geodesics, which are
Euclidean half-circles, hitting the real axis at $\nu_0$, at a
right angle.

In addition to (\ref{equation24}) they are general solutions
\cite{Terras} of the form
\begin{equation}
\phi_s(z)=\sum_{\gamma \in \Gamma_{\infty}/\Gamma} (\Im
\gamma(z))^s, ~~s=\frac{1}{2}+I k,
 \label{equation30}
\end{equation}
where $\Gamma_{\infty}$ is the stabilizer in the modular group
$\Gamma$ of the cusp at inf\mbox{}inity, i.e. the subgroup of
integer translations $z \rightarrow z+n$. The solution corresponds
to waves scattered again and again from the modular surface
$\it{F}$. For arbitrary $s$ the series (\ref{equation30}) can be
rewritten as
$\phi_s(z)=\frac{1}{2}y^s\sum_{(p_i,q_i)=1}\frac{1}{|p_iz+q'_i|^{2s}}$
\cite{Terras}, with the summation performed over all coprime
numbers $(p_i,q_i)=1$. The convergence is ensured for $\Re(s)>1$.

The series (\ref{equation30}) satisf\mbox{}ies a functional
equation
\begin{equation}
\xi(2s)\phi_s(z)=\xi(2-2s)\phi_{1-s}(z),
 \label{equation32}
\end{equation}
where $\xi(s)=\pi^{-s/2}\Gamma(s/2)\zeta(s)$ is the completed
Riemann zeta function. It follows that (\ref{equation30}) is
expanded as
\begin{eqnarray}
&\phi_s(z)=y^s+S(s)y^{1-s}+T_s(y),\nonumber \\
&\mbox{with}~~S(s)=A(s)Z(s), \nonumber\\
&Z(s)=\frac{\zeta(s-1)}{\zeta(s)},~~A(s)=\frac{\Gamma(1/2)\Gamma(s-1/2)}{\Gamma(s)},
 \label{equation33}
\end{eqnarray}
and the remaining term vanishes exponentially when $y\rightarrow
\infty$. For $s=\frac{1}{2}+Ik$, that is $s$ on the critical line,
the remainder is $T_s(y)=0$. In such a case the scattering
coeff\mbox{}icient equals
\begin{equation}
S(\frac{1}{2}+Ik)=\frac{\xi(2Ik)}{\xi(1+2Ik)},~~\mbox{with}~|S(\frac{1}{2}+I
k)|=1,
\end{equation}
that is the f\mbox{}lux of the ref\mbox{}lected wave is equal and
opposite sign the incoming f\mbox{}lux. The phase angle
represented on Fig. 4 is a very complicated function of $k$ and is
even considered as a prototype for quantum chaos
\cite{Gutzwiller}.

It is observed that the term $A(\frac{1}{2}+Ik)$ in
(\ref{equation33}) increases while $|Z(\frac{1}{2}+Ik)|$ decreases
monotonously to preserve the modulus 1 of $S(\frac{1}{2}+Ik)$. All
the stochastic dynamics is encoded in the function
$Z(\frac{1}{2}+Ik)$. Fig. 4 represents the phase angle $\theta(k)$
in the scattering coeff\mbox{}icient $S(\frac{1}{2}+I k)=\exp(I
\theta(k))$ in comparison to the one $\theta_0(k)$ of
$Z(\frac{1}{2}+I k)$. Apart for weak changes the two curves looks
similar; this is conf\mbox{}irmed from the power spectral density
which is about the same for the two cases.
%
%

Looking at the logarithmic derivative of the scattering
coeff\mbox{}icient $S(s)=\exp(i\theta(s))$, one can get the
counting function
\begin{equation}
\theta'(k)=\frac{d\ln S(s)}{ds}~~\mbox{at}~s=\frac{1}{2}+Ik.
\end{equation}
which is related to the stochastic factor $Z(s)$ as
\begin{equation}
-\frac{Z'(s)}{Z(s)}=\sum_{n=1}^{\infty}\frac{b(n)}{n^s}
=s\int_1^{\infty}t^{-s-1}B(t)dt, \label{equation34}
\end{equation}
with $B(t)=\sum_{n=1}^{\infty}b(n)$, $b(n)=\Lambda(n)\phi(n)/n$.
By inverting the Mellin transform in (\ref{equation33}) one gets

\begin{equation}
B(t)=\frac{1}{2I\pi}\int_{\Re(s)-I\infty}^{\Re(s)+I\infty}-\frac{Z'(s)}{Z(s)}.\frac{t^s}{s}ds.
\end{equation}
This extends the calculation performed in (\ref{Riemann}) for
getting the error term in the summatory Mangoldt function
$\psi(t)$. There the Riemann zeta function $\zeta(s)$ replaces the
quotient $Z(s)=\zeta(s)/\zeta(s+1)$ given in (\ref{equation11}).
As reminded in Fig. 2, the error term in the summatory
modif\mbox{}ied Mangoldt function is very close to a
$1/f^{2\alpha}$ noise, with $\alpha$ the Golden ratio.

\section{Concluding Remarks}
From its def\mbox{}inition $1/f$ noise is attached to the use of
the fast Fourier transform (FFT). But the FFT refers to the fast
calculation of the discrete Fourier transform (DFT) with a
f\mbox{}inite period $q=2^l$, $l$ a positive integer. In the DFT
one starts with all $q^{\rm{th}}$ roots of the unity $\exp(2i\pi
p/q)$, $p=1\ldots q $ and the signal analysis of the arithmetical
sequence $x(n)$  is performed by projecting onto the $n^{\rm{th}}$
powers (or characters of \textit{Z}/q\textit{Z}) with well known
formulas.

The signal analysis based on the DFT is not well suited to
aperiodic sequences with many resonances (by nature a resonance is
a primitive root of the unity: $(p,q)=1$), and the FFT may fail to
discover the underlying structure in the spectrum. We recently
introduced a new method based on Ramanujan sums
\cite{PRE02},\cite{Hardy}.
\begin{equation}
c_q(n)=\sum_ {p=1}^q \exp(2i\pi \frac{p}{q} n)
~~\mbox{with}~(p,q)=1,
\end{equation}
which are $n^{\rm{th}}$ powers of the $q^{\rm{th}}$ primitive
roots of the unity. The sums are evaluated from the use of
M\"{o}bius transforms as $c_q(n)=\mu\left(
\frac{q}{(q,n)}\right)\frac{\phi(q)}{\phi\left(
\frac{q}{(q,n)}\right)}$. Here $(q,n)$ is the greatest common
divisor $(q,n)$ of $q$ and $n$ and $\mu(n)$ is the M\"{o}bius
function. It is $0$ if $n$ contains a square, $1$ if $n=1$ and
$(-1)^k$ if $n$ is the product of $k$ distinct primes.

The sums are quasiperiodic versus the time $n$
($c_1=\overline{1};~ c_2=\overline{-1,1};~
c_3=\overline{-1,-1,2}$, where the bar indicates the period; for
example $c_3(4)=-1$) and aperiodic versus the order $q$ of the
resonance. In particular $c_q(n)=\mu(q)$ whenever $(n,q)=1$.

M\"{o}bius function can be considered as a coding sequence for
prime numbers, as it is the case of Mangoldt function.  Mangoldt
function is related to M\"{o}bius function thanks to the Ramanujan
sums expansion found by Hardy \cite{Gadiyar}
\begin{equation}
b(n)=\frac{\phi(n)}{n}\Lambda(n)=\sum_{q=1}^{\infty}\frac{\mu(q)}{\phi(q)}c_q(n).
\label{equab}
\end{equation}
We call such a type of Fourier expansion a Ramanujan-Fourier
transform (RFT). General formulas are given in our recent
publication \cite{PRE02} and in the paper by Gadiyar
\cite{Gadiyar}. This author also reports on a stimulating
conjecture relating the autocorrelation function of $b(n)$ and the
problem of prime pairs. In the special case (\ref{equab}), it is
clear that $\mu(q)/\phi(q)$ is the RFT of the modif\mbox{}ied
Mangoldt sequence $b(n)$.
\begin{figure}[htb]
\centerline{
\includegraphics[width=6.5cm,angle=-90]{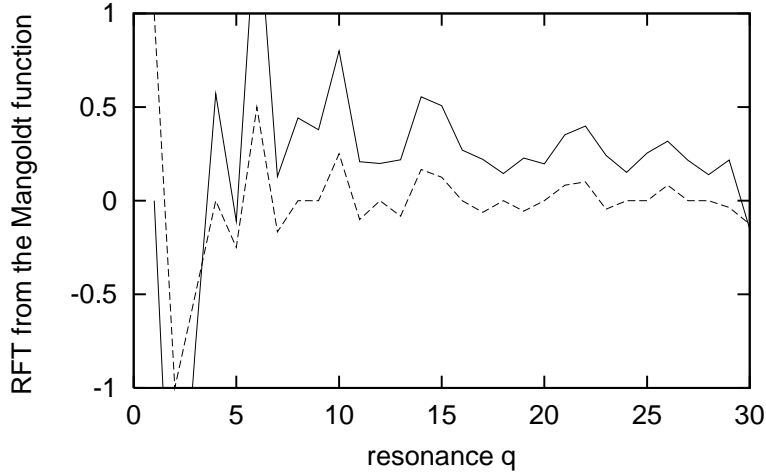}}
\caption{Ramanujan-Fourier transform (RFT) of the error term
(upper curve) of modif\mbox{}ied Mangoldt function $b(n)$  in
comparison to the function $\mu(q)/\phi(q)$(lower curve).}
\end{figure}

Using Ramanujan-Fourier analysis the $1/f^{2\alpha}$ power
spectrum gets replaced by a new signature shown on Fig. 5, very
close to $\mu(q)/\phi(q)$ (up to a scaling factor). There is thus
a deep relationship between $1/f$ noise in phase locking, the
Golden ratio, the M\"{o}bius function, the modif\mbox{}ied
Mangoldt function, the frequency of windings around the modular
surface, and the Riemann hypothesis. All these ingredients arise
from the hyperbolic geometry of the half-plane.

\end{document}